\documentstyle[prd,aps,preprint,psfig]{revtex}

\newcommand{\CK}{{\cal K}}
\newcommand{\CL}{{\cal L}}

\newcommand{\CR}{{\cal R}}

\newcommand{\bear}{\begin{array}}  \newcommand{\eear}{\end{array}}
\newcommand{\bea}{\begin{eqnarray}}  \newcommand{\eea}{\end{eqnarray}}
\newcommand{\beq}{\begin{equation}}  \newcommand{\eeq}{\end{equation}}
\newcommand{\bef}{\begin{figure}}  \newcommand{\eef}{\end{figure}}
\newcommand{\bec}{\begin{center}}  \newcommand{\eec}{\end{center}}
\newcommand{\non}{\nonumber}  
\newcommand{\lmk}{\left(}  \newcommand{\rmk}{\right)}
\newcommand{\lkk}{\left[}  \newcommand{\rkk}{\right]}
\newcommand{\lhk}{\left \{ }  
\newcommand{\del}{\partial}  

\newcommand{\bib}{\bibitem} 
\newcommand{\la}{\left\langle} \newcommand{\ra}{\right\rangle}
 
\newcommand{\gtilde} {~ \raisebox{-1ex}{$\stackrel{\textstyle >}{\sim}$} ~}

\def\IB#1#2#3{{\bf #1}, #2 (19#3)}
\def\IBID#1#2#3{{\it ibid}. {\bf #1}, #2 (19#3)}
\def\NPB#1#2#3{Nucl. Phys. {\bf B#1}, #2 (19#3)}

\def\PLB#1#2#3{Phys. Lett. B {\bf #1}, #2 (19#3)}
\def\PLBold#1#2#3{Phys. Lett. {\bf#1B}, #2 (19#3)}

\def\PRD#1#2#3{Phys. Rev. D {\bf #1}, #2 (19#3)}
\def\PRDD#1#2#3{Phys. Rev. D {\bf #1}, #2 (20#3)}

\def\PRL#1#2#3{Phys. Rev. Lett. {\bf#1}, #2 (19#3)}

\def\APJL#1#2#3{Astrophys. J. Lett. {\bf #1}, L#2 (19#3)}

\def\PTP#1#2#3{Prog. Theor. Phys. {\bf #1}, #2 (19#3)}

\def\JP#1#2#3{J. Phys. A {\bf #1}, #2 (19#3)}
\def\JHEP#1#2#3{J. High Energy Phys. {\bf #1}, #2 (19#3)}

\begin{document}

%@twocolumnfalse\endcsname
%%
%%
\tighten
\draft
\title{Topological Inflation in Supergravity}
\author{M. Kawasaki}
\address{Research Center for the Early Universe, University of Tokyo,
  Tokyo 113-0033, Japan}
\author{Nobuyuki Sakai}
\address{DAMTP, Centre for Mathematical Sciences, University of
Cambridge, Wilberforce Road, Cambridge CB3 0WA, United Kingdom}  
\author{Masahide Yamaguchi}
\address{Research Center for the Early Universe, University of Tokyo,
  Tokyo 113-0033, Japan}
\author{T. Yanagida}
\address{Department of Physics, University of Tokyo, Tokyo 113-0033,
Japan \\ and \\ Research Center for the Early Universe, University of
Tokyo, Tokyo 113-0033, Japan}
%%%

\date{\today}

\maketitle

\begin{abstract}

    We investigate a topological inflation model in supergravity. By
    means of numerical simulations, it is confirmed that topological
    inflation can take place in supergravity. We also show that the
    condition for successful inflation depends not only on the
    vacuum-expectation value (VEV) of inflaton field but also on the
    form of its K\"ahler potential. In fact, it is found that the
    required VEV of the inflaton $\varphi$ can be as small as $\langle
    \varphi \rangle \simeq 1 \times M_G$, where $M_{G}$ is the
    gravitational scale.

\end{abstract}


%]

%%%%%%%%%%%%%%%%%%%%%%%%%%%%%%%%%%%%%%%%%%%%%%%%%%%%%%%%%%%%%%%%%%%%%%
\section{Introduction}
%%%%%%%%%%%%%%%%%%%%%%%%%%%%%%%%%%%%%%%%%%%%%%%%%%%%%%%%%%%%%%%%%%%%%%

Superstring theories compactified on $(3+1)$-dimensional space-time
have many discrete symmetries in the low-energy effective Lagrangian
\cite{discrete}. A spontaneous breakdown of such discrete symmetries
creates topological defects, i.e. domain walls, in the early universe
\cite{KIB}. If the vacuum-expectation value (VEV) of a scalar field
$\varphi$ is larger than the gravitational scale $M_{G} \simeq 2
\times 10^{18}$~GeV, the region inside the wall undergoes inflationary
expansion and eventually becomes the present whole universe
\cite{Linde,Vilenkin}. If the universe is open at the beginning, it
expands and the spontaneous breakdown of the symmetries always takes
place at some epoch in the early universe. It has been recently argued
that the quantum creation of the open universe may take place with
appropriate continuation from the Euclidean instanton \cite{open}.
Thus, topological inflation is a natural consequence of the dynamics
of the system, and it does not require any fine-tuning of initial
conditions for the beginning universe. Furthermore, it does not cause
the ``graceful exit" problem and the universe becomes homogeneously
radiation dominated after reheating.

A simple and interesting model for topological inflation was proposed
in the framework of supergravity \cite{Izawa}.\footnote{Other
topological inflation models were studied in the superstring inspired
models \cite{BBN,EKOY}.} However, it was not explicitly shown whether
topological inflation really takes place. In this paper, we perform a
numerical analysis on the above model and show that topological
inflation indeed occurs in a wide range of parameter space. We also
show that the condition for successful inflation depends not only on
the superpotential, which determines the vacuum expectation value
(VEV) of inflaton $\varphi$, but also on the form of its K\"ahler
potential.  We in fact find that the required VEV can be as small as
$\langle \varphi \rangle \simeq 1 \times M_G$, which is far below the
lower bound of $\la \varphi \ra = \eta_{cr} \simeq 1.7M_{G}$ derived
in Ref. \cite{sakai}.

%%%%%%%%%%%%%%%%%%%%%%%%%%%%%%%%%%%%%%%%%%%%%%%%%%%%%%%%%%%%%%%%%%%%%%
\section{Topological Inflation Model}
%%%%%%%%%%%%%%%%%%%%%%%%%%%%%%%%%%%%%%%%%%%%%%%%%%%%%%%%%%%%%%%%%%%%%%

We begin with the topological inflation model proposed in
Ref.~\cite{Izawa}, which is based on $R$-invariant supergravity. The
gravitational scale $M_{G}$ is set to be unity below. In this model
the superpotential for the inflaton superfield $\phi(x,\theta)$ is
given by
\beq
  W = v^2 X (1 - g \phi^2).
  \label{eq:super-pot}
\eeq
Here, we have imposed $U(1)_R \times Z_{2}$ symmetry and omitted
higher-order terms for simplicity. Under the $U(1)_{R}$ we assume
\beq
  X(\theta)  \rightarrow  e^{-2i\alpha} X(\theta e^{i\alpha}),~~~
    \phi(\theta)  \rightarrow \phi(\theta e^{i\alpha}).
\eeq
We also assume that the superfield $X$ is even and $\phi$ is odd under
the $Z_2$.  This discrete $Z_2$ symmetry is an essential ingredient
for the topological inflation~\cite{Linde,Vilenkin}. In the above
superpotential (\ref{eq:super-pot}), we always take $v^{2}$ and $g$
to be real constants without loss of generality.

The $R$- and $Z_2$-invariant K\"ahler potential is given by
\beq
    \label{eq:kpot}
    K(\phi,X) = |X|^2 + |\phi|^2 + k_{1}|X|^2|\phi|^2 
    + \frac{k_{2}}{4}|X|^4 +\frac{k_{3}}{4}|\phi|^4
    + \cdots ,
\eeq
where $k_{1}$, $k_{2}$ and $k_{3}$ are constants of order unity.

The potential of a scalar component of the superfields $X(x,\theta)$
and $\phi(x,\theta)$ in supergravity is given by
\beq
  V = e^{K} \left\{ \left(
      \frac{\partial^2K}{\partial z_{i}\partial z_{j}^{*}}
    \right)^{-1}D_{z_{i}}W D_{z_{j}^{*}}W^{*}
    - 3 |W|^{2}\right\}~~~~~~~(z_i = \phi, X ),
  \label{eq:potential}
\eeq
with 
\beq
  D_{z_i}W = \frac{\partial W}{\partial z_{i}} 
    + \frac{\partial K}{\partial z_{i}}W.
  \label{eq:DW}
\eeq
This potential yields an $R$-invariant vacuum 
\bea
  \langle X \rangle = 0,
    \quad
     \langle \phi \rangle = 
    \frac{1}{\sqrt{g}} \equiv \frac{\eta}{\sqrt{2}},
  \label{eq:eta-g}
\eea
at which the potential energy vanishes. Here, the scalar components of
the superfields are denoted by the same symbols as the corresponding
superfields. If $\eta$ is larger than the critical value $\eta_{\rm
  cr}$ which will be discussed in the next section, the topological
inflation occurs.

For $|X|$ and $|\phi| \ll 1$, we approximately rewrite the
potential~(\ref{eq:potential}) as
\beq
  V \simeq v^4|1 - g\phi^{2}|^{2} + v^4 (1 - k_1)|\phi|^2 
  - k_2v^4 |X|^2.
  \label{eq:eff-pot}
\eeq
If $k_2 \lesssim -1$, $X$ field quickly settles down to the origin and
we set $X =0$ in our analysis taking $k_{2} \lesssim -1$.  For $g >
0$, we can identify the inflaton field $\varphi(x)/\sqrt{2}$ with the
real part of the field $\phi(x)$ since the imaginary part of $\phi(x)$
has a positive mass and the real part has a negative mass.  Because
the positive mass of the imaginary part is larger than the size of the
negative mass, the imaginary part is irrelevant for the inflation
dynamics and hence we neglect it. Then, we obtain a potential for the
inflaton for $\varphi \ll 1$:
\beq
  V(\varphi) \simeq v^4 - \frac{\kappa}{2}v^4\varphi^2,
  \label{eq:eff-pot2}
\eeq
where
\beq
  \kappa \equiv 2g + k_{1} - 1.
  \label{eq:kappa}
\eeq
The slow-roll condition for the inflaton $\varphi$ is satisfied for
$0< \kappa < 1$ and $0 \lesssim \varphi \lesssim 1$.\footnote{We can
always take $\varphi$ positive since we have the $Z_{2}$ symmetry
$(\phi \rightarrow - \phi)$.} The Hubble parameter during the
inflation is given by $H \simeq v^2/\sqrt{3}$.  The scale factor of
the universe increases by a factor of $e^N$ when the inflaton
$\varphi$ rolls slowly down the potential from $\varphi_N$ to $1$.
The $e$-fold number $N$ is given by
\bea
  N \simeq - {1 \over \kappa}\ln{\varphi_N}.
  \label{eq:N-efold}
\eea

The amplitude of primordial density fluctuations $\delta \rho/\rho$ 
due to this inflation is written as 
\beq
  \frac{\delta\rho}{\rho} \simeq \frac{1}{5\sqrt{3}\pi} 
  \frac{v^{2}}{\kappa\varphi_{N}} \sim 2.0 \times 10^{-5}.
  \label{eq:density}
\eeq
The normalization is given by the data of anisotropies of the cosmic
microwave background radiation (CMB) by the COBE satellite
\cite{COBE}. Since the $e$-fold number $N$ corresponding to the COBE
scale is about $60$, which leads to
\bea
    v  \simeq  2.3 \times 10^{-2} \sqrt{\kappa} 
    e^{-\frac{\kappa N}{2}}|_{N=60}
    \simeq 1.8 \times 10^{-3} - 3.6 \times 10^{-4},
\eea
for $0.02 \le \kappa \le 0.1$.

The interesting point on the above density fluctuations is that it
results in the tilted spectrum whose spectrum index $n_s$ is given by
\beq
    \label{eq:new-index}
    n_s \simeq 1 - 2 \kappa.
\eeq
We may expect a possible deviation from the Harrison-Zeldvich
scale-invariant spectrum $n_s = 1$. Observational constraint on $n_s$
is $|n_s-1|< 0.2$~\cite{COBE}, which implies $0 < \kappa < 0.1$.

After inflation ends, the inflaton $\varphi$ may decay into ordinary
particles as discussed in Ref.~\cite{Izawa} and the reheating
temperature is low enough to avoid overproduction of
gravitinos~\cite{Ellis} which are thermally produced at the reheating
epoch. Recently, nonthermal production at the preheating stage was
found to be important in some inflation models~\cite{Kallosh}. For the
present model, non-thermal production of gravitinos at the preheating
phase is roughly estimated as
\bea
  \lmk \frac{n_{3/2}}{s} \rmk_{\rm non-TH} 
   \sim \, \frac{m_{\varphi}^{3}}{v^{4}/T_{R}} \,
    \lesssim 10^{-14} \lmk \frac{T_R}{10^{10}{\rm GeV}} \rmk,
\eea
where $m_{\varphi}(\simeq v^{2})$, $n_{3/2}$, $s$, and $T_R$ are the
mass of the inflaton, the number density of gravitinos, entropy
density and reheating temperature, respectively. This is much less
than the thermal production given by $(n_{3/2}/s)_{\rm TH} \sim
10^{-11}(T_R/10^{10}{\rm GeV})$ and hence we can neglect the
nonthermal production of gravitinos.

%%%%%%%%%%%%%%%%%%%%%%%%%%%%%%%%%%%%%%%%%%%%%%%%%%%%%%%%%%%%%%%%%%%%%%
\section{Numerical Simulation}
%%%%%%%%%%%%%%%%%%%%%%%%%%%%%%%%%%%%%%%%%%%%%%%%%%%%%%%%%%%%%%%%%%%%%%

We perform numerical simulations to decide whether topological
inflation takes place in supergravity and determine the condition for
successful topological inflation. For the purpose, we follow time
evolution of the domain wall and investigate whether it inflates or
not. Since we consider a planar domain wall whose width is order of
the horizon scale, we cannot adopt the Friedmann Robertson-Walker
metric. Instead, we assume that the spacetime has a reflection
symmetry of the coordinate $x$ perpendicular to the wall and adopt the
metric given by
\beq
  ds^{2} = -~dt^{2} + A^{2}(t,|x|)~dx^{2} 
                    + B^{2}(t,|x|)~(dy^{2} + dz^{2}),
  \label{eq:metric}                  
\eeq
where $A$ and $B$ correspond to the scale factors in the direction of
$x$ and $y$-$z$, respectively. If the inflation occurs and the proper
width of the wall becomes much larger than the horizon scale, $A$ and
$B$ expand as $A \sim B \propto e^{Ht}$ (as shown later) and the
universe approaches the de Sitter spacetime.  Thus, we examine whether
the proper width of the wall becomes much larger than the horizon
scale. Once it is realized, the universe expands exponentially
\cite{Linde,Vilenkin}.

We adopt the numerical technique developed in Ref.~\cite{sakai}. The
Einstein-Hilbert action is given by
\beq
  S = \int d^{4}x \sqrt{-g} \lkk \frac12 \CR 
         - \frac12 (\del_{\mu}\varphi)^{2} - V(\varphi) \rkk.
  \label{eq:action}
\eeq
Variating the above action with respect to the metric $g_{\mu\nu}$, we
obtain the Einstein equations,
\beq
  G_{\mu\nu} \equiv \CR_{\mu\nu} - \frac12 \CR = T_{\mu\nu},
\eeq
where $T_{\mu\nu}$ is the energy-momentum tensor,
\beq
  T_{\mu\nu} = \del_{\mu}\varphi \del_{\nu}\varphi 
               - g_{\mu\nu} \lkk 
                   \frac12(\del_{\mu}\varphi)^2 + V(\varphi) 
                            \rkk.
\eeq
Variation with respect to the scalar field $\varphi$ gives the
equation of motion for the scalar field $\varphi$,
\beq
  \Box \varphi = \frac{d V(\varphi)}{d \varphi}.
\eeq
In order to make it easier to follow time evolution of the system, we
choose certain combinations of Einstein equations, which read
\bea
  -G^{0}_{0} &=& \CK^{2}_{2}(2\CK-3\CK^{2}_{2}) - \frac{2B''}{A^{2}B}
                 - \frac{B'^{2}}{A^{2}B^{2}} + \frac{2A'B'}{A^{3}B}
                 \non \\
             &=& \frac{\dot\varphi^{2}}{2} 
                 + \frac{\varphi'^{2}}{2A^{2}} + V(\varphi), 
  \label{eq:einstein1} \\
  \frac12 G_{01} &=& {\CK^{2}_{2}}' + \frac{B'}{B}(3\CK^{2}_{2}-\CK) \non \\
                 &=& \frac12\dot\varphi\varphi', 
  \label{eq:einstein2} \\
  \frac12(G^{1}_{1} + G^{2}_{2} + G^{3}_{3} - G^{0}_{0})
             &=& \dot{\CK} - (\CK^{1}_{1})^{2} - 2(\CK^{2}_{2})^{2} \non \\
             &=& \dot\varphi^{2} - V(\varphi), 
  \label{eq:einstein3} \\
  -\CR^{2}_{2} - \frac12 G^{0}_{0} 
             &=& \dot{\CK^{2}_{2}} + \frac{{B'}^2}{2A^{2}B^{2}} 
                 - \frac32 (\CK^{2}_{2})^{2} \non \\
             &=& \frac{\dot\varphi^{2}}{4} 
                 + \frac{\varphi'^{2}}{4A^{2}} - \frac{V(\varphi)}{2},
  \label{eq:einstein4}
\eea
where an overdot denotes the time derivative and a dash the spatial
derivative. $\CK_{ij}$ are the extrinsic curvature tensors of constant
time hypersurface, given by
\beq
  \CK^{1}_{1} = - \frac{\dot A}{A}, \qquad
  \CK^{2}_{2} = \CK^{3}_{3} = - \frac{\dot B}{B},
  \label{eq:extrinsic}
\eeq
and $\CK$ denotes its trace $\CK \equiv \CK^{i}_{i}$. The equation of
motion for the scalar field $\varphi$ becomes
\beq
  \ddot\varphi - \CK\dot\varphi - \frac{\varphi''}{A^{2}} 
    - \lmk -\frac{A'}{A} + \frac{2B'}{B} \rmk \frac{\varphi'}{A^{2}}
    + \frac{dV(\varphi)}{d\varphi} = 0.
  \label{eq:eom} 
\eeq

We set the initial condition for numerical simulations. First, we
consider an initial configuration of a domain wall. For the
convenience of numerical calculations, we take only the region
$|x|/\delta \leq 2$ \footnote{We have confirmed that the results do
not change even if we take wider ranges of the direction $x$ (e.g.
$|x|/\delta = 3$).} where $\delta(=\frac{\eta}{\sqrt2 v^{2}})$ is the
width of the domain wall and $\eta$ is the VEV of $\varphi$. We impose
the free boundary condition, that is, $\varphi' = A' = B' = 0$ at the
boundaries $x/\delta = -2$ and $x/\delta = 2$. Then, we adopt the
following initial configuration for the domain wall so that the
gradient of the field disappears at the boundaries,
\beq
  \varphi(t = 0, x) = \lhk
     \bear{ll}
       \displaystyle{
       \eta \lkk 
          \frac{x}{\delta} 
          - \frac54 \lmk\frac{8}{15}\frac{x}{\delta} \rmk^{3}    
          + \frac38 \lmk\frac{8}{15}\frac{x}{\delta} \rmk^{5}    
            \rkk} \qquad
       \lmk 0 \leq \frac{x}{\delta} \leq \frac{15}{8}
       \rmk, \\ [0.5cm] 
       \displaystyle{\eta} \hspace{6.4cm}
       \lmk \frac{15}{8} \leq \frac{x}{\delta} \leq 2 \rmk,
     \eear
     \right.
   \label{eq:initconfig}
\eeq
with $\varphi(t=0,-x) = - \varphi(t=0, x)$ for $-2 \leq x/\delta \leq
0$. This is a deformed version of the static domain wall solution in a
flat spacetime, $\varphi_{\rm flat} = \eta\tanh(x/\delta)$. The
function $\varphi(t=0,x)$ is decided so as to satisfy the following
three conditions: (1)~it is a fifth-order odd polynomial function of
$x$, (2)~the first term coincides with that of the expansion of
$\varphi_{\rm flat}$, (3)~it is smooth at $x/\delta = 15/8$, that is,
$\varphi' = \varphi'' = 0$ at $x/\delta = 15/8$. Also, $\dot\varphi$
is set to be 0.

Next, on the initial hypersurface, we determine $A, B, \CK_{2}^{2},
\CK$ so as to satisfy the Hamiltonian constraint (\ref{eq:einstein1})
and the momentum constraint (\ref{eq:einstein2}). We have freedom for
the initial hypersurface to have homogeneous and isotropic curvature,
which automatically satisfies the momentum constraint
(\ref{eq:einstein2}). This choice leads to
\beq
  \frac{\CK}{3} = \CK^{1}_{1} = \CK^{2}_{2} = 
      \textrm{``negative''~const},
  \label{eq:initK}
\eeq
where ``negative'' implies that the universe is in an expanding phase.
Furthermore, we can take the conformally flat spatial gauge, $A = B$, on
the initial hypersurface and set $A = B = 1$ at $x = 0$. Finally, we
determine the negative value of $\CK$. Since we adopt the reflection
symmetry of the coordinate $x$ and the free boundary condition, the
condition $A' = B' = 0$ at $x = 0$ and $x = \pm2\delta$ must be
satisfied. $\CK$ is determined so that the Hamiltonian constraint
(\ref{eq:einstein1}) satisfies the above conditions.

Now the initial settings are completed and hence we have only to
follow the time evolution of five variables, $A, B, \CK, \CK_{2}^{2}$,
and $\varphi$. Note that we have introduced five variables, $A, B,
\CK, \CK_{2}^{2}$, and $\varphi$ though only three variables are
independent.  This is partly because the second order differential
equations have been reduced to the first order differential equations.
Moreover, as for time evolution of $\CK_{2}^{2}$, we use
Eq.(\ref{eq:einstein4}) only at $x = 0$ and acquire the value at $x
\ne 0$ by integrating Eq.(\ref{eq:einstein2}) in the direction of $x$
in order to avoid numerical instability.

When inflation takes place, $A$ and $B$ grow exponentially so that the
proper distance from the domain wall core ($x = 0$) also increases
exponentially. In order to see whether this happens or not, we follow
time evolution of the width of the wall for a given potential
$V(\varphi)$.

To fix the potential $V(\varphi)$, we first consider the K\"ahler
potential with only terms up to the fourth order,
\beq
  K(\phi,X) = |X|^2 + |\phi|^2 + k_{1}|X|^2|\phi|^2 
        +\frac{k_{2}}{4}|X|^4 +\frac{k_{3}}{4}|\phi|^4.
  \label{eq:kpotfour}
\eeq
Then, the Lagrangian density is given by
\bea
  \CL(\phi) &=& \CL_{kin}(\phi) - V(\phi) \non \\
      &=& - (1 + k_{3}|\phi|^{2})\del_{\mu}\phi\del^{\mu}\phi^{\ast}
          - v^{4} 
            |1 - g\phi^{2}|^{2}~
            \frac{\exp \lmk |\phi|^{2} + \frac{k_{3}}{4} |\phi^{4}| \rmk}
            {1 + k_{1}|\phi|^{2}}.
  \label{eq:complexlag}   
\eea 
Here we have set $X = 0$. Identifying the inflaton field
$\varphi(x)/\sqrt{2}$ with the real part of the field $\phi(x)$, the
Lagrangian density becomes
\bea
    \CL(\varphi) &=& \CL_{kin}(\varphi) - V(\varphi) \non \\
      &=& - \frac12\left(1 + \frac12 k_{3}\varphi^{2}\right) 
            (\del_{\mu}\varphi)^{2}- v^{4} 
            \left(1 - \frac{g}{2}\varphi^{2}\right)^{2}~
             \frac{\exp \lmk 
                 \frac12 \varphi^{2} + \frac{k_{3}}{16} \varphi^{4} 
                 \rmk}
            {1 + \frac{k_{1}}{2}\varphi^{2}},
  \label{eq:reallag}   
\eea 
with the VEV $\eta = \sqrt{2/g}$. In the present model we have four
free parameters $k_1, k_2, k_3$ and $g$. However, $k_2 (\lesssim -1) $
only works as a stabilizer of the $X$ field as explained before and it
is not important for the dynamics of topological inflation itself.
Once the $X$ field is stabilized at $X = 0$, the potential
$V(\varphi)$ does not depend on $k_2$. $k_3$ is almost irrelevant for
the dynamics of $\phi$ field and only changes its VEV due to the
redefinition of $\varphi$ with a canonical kinetic term. Then, we set
$k_{3} = 0$ first and later consider the case of nonzero $k_{3}$.
Thus, we have only two relevant parameters, $k_1$ and $g$. The
potential $V(\varphi)$ has a pole at $\varphi = \sqrt{2/|k_{1}|}$ for
$k_{1} < 0$. But, we are only interested in the dynamics of $\varphi$
up to the VEV $\eta$ so that there is no problem if $|k_{1}| < g$ for
$k_{1} < 0$.

We introduce dimensionless quantities, $\bar\varphi = \varphi/M_{G},
\bar x = x H(x = 0), \bar t = t H(x = 0), \bar \CK_{ij} = \CK_{ij} /
H(x = 0)$, and $\bar\delta = \delta H(x = 0) = 1 / (\sqrt{3g})$ where
$H(x = 0) = v^{2}/\sqrt{3}$. As the first step, we consider the
simplest case of $k_{1} = 0$ and $g = 0.5~($i.e., $\eta = 2.0)$, which
leads to the spectral index $n = 1$. Time evolution of the domain wall
is depicted in Fig.  \ref{fig:minimal}. The vertical axis represents
the value of the scalar field. The horizontal axis represents the
proper distance from the domain wall core. As time elapses, the domain
wall (the region for $\varphi \lesssim 0.8$) expands and topological
inflation really takes place.  As shown in Figs.  \ref{fig:tk1k2} and
\ref{fig:k1k2x0}, once the domain wall expands enough, the scale
factors $A$ and $B$ increase at the same rate, that is, $-\CK^{1}_{1}
\sim -\CK^{2}_{2} \sim H(x = 0)$ inside the wall. Thus, the universe
expands exponentially and approaches the de Sitter spacetime.

We consider the dependence on $\kappa$ for the VEV fixed, $\eta =
2.0~(g = 0.5)$. The result is depicted in Fig. \ref{fig:g=0.5v=1.8-3}.
For large $\kappa$, the domain wall cannot inflate enough. $\kappa$
controls the mass scale of the scalar field $\varphi$ near the origin.
As $\kappa$ becomes larger, the scalar field $\varphi$ rolls down
faster so that only the small region of the original domain wall earns
the vacuum energy and cannot overcome the gradient energy.

The dependence on $v$ is studied, which determines the energy scale of
the domain wall. The result with the same parameters in Fig.
\ref{fig:g=0.5v=1.8-3} except for $v = 3.6 \times 10^{-4}$ is shown in
Fig. \ref{fig:g=0.5v=3.6-4}. The results are quite the same and have
no dependence on $v$. This can be interpreted as follows: First, the
dependence on $v$ only appears through the Hubble parameter during the
inflation given by $H \simeq v^2/\sqrt{3}$. But, since the width of
the domain wall $\delta \sim 1 / (\sqrt{3g} H)$, the rough criterion
$\delta > H^{-1}$ becomes independent of $H$, that is, $v$.  Also,
rewriting the potential as $V(\varphi) \sim \lambda
(\varphi^{2}-\eta^{2})$ with $\lambda = (v/\eta)^{4}$, the
independence of $v$ is equivalent to that of $\lambda$. The
independence of $v$ is also found in Ref.~\cite{sakai}.

We now consider the cases of the nonzero $k_{3}$. Since the kinetic
term $\CL_{kin}$ is not canonical in these cases, we define the scalar
field $\Phi$ with the canonical kinetic term as
\beq
   \Phi = \int_{0}^{\varphi} d\varphi \sqrt{1+\frac12 k_{3}\varphi^{2}}  
                     = \lhk
     \bear{ll}
       \displaystyle{
         \frac12\varphi \sqrt{1+\frac12 k_{3}\varphi^{2}} 
         + \frac{{\rm arcsinh} \lmk \sqrt{\frac{k_{3}}{2}} \varphi \rmk}
                {\sqrt{2k_{3}}} } \qquad \qquad
       \lmk k_{3} \geq 0\rmk, \\ [0.5cm]
        \displaystyle{
         \frac12\varphi \sqrt{1-\frac12 |k_{3}|\varphi^{2}} 
         + \frac{{\rm arcsin} \lmk \sqrt{\frac{|k_{3}|}{2}} \varphi \rmk}
                {\sqrt{2|k_{3}|}} }
         \hspace{1.4cm}
       \lmk k_{3} < 0 \rmk.
     \eear
     \right.
   \label{eq:k_{3}} 
\eeq
The results for the same parameters in Fig. \ref{fig:minimal} except
for $k_{3}=\pm 0.3$ are depicted in Fig. \ref{fig:t_k3}. The positive
$k_{3}$ encourages the occurrence of topological inflation while the
negative $k_{3}$ discourages. This is because the VEV of $\Phi$ is
larger than $\eta$ for the positive $k_{3}$ while smaller for the
negative $k_{3}$.

Finally, we discuss the criterion for the VEV of $\varphi$ for
successful topological inflation within $0 \leq \kappa \lesssim 0.1$.
In previous analyses, we search for only the parameter region
satisfying $|k_{1}| < g$ for negative $k_{1}$ because of the
appearance of a pole of the potential. The condition $|k_{1}| < g$
leads to $g < 1 + \kappa$ so that $\eta > 1.41(1.35)$ for $\kappa =
0.0(0.1)$, where we have confirmed the occurrence of topological
inflation. In order to obtain the lower limit of $\eta$, we add to the
K\"ahler potential the sixth order terms,
\beq
   \Delta K =
      l_{1}|X|^{2}|\phi|^{4} + l_{2}|X|^{4}|\phi|^{2} + 
      \frac{l_{3}}{9} |X|^{6} + \frac{l_{4}}{9} |\phi|^{6},
\eeq
where there are four parameters but only $l_{1}$ is relevant.  If we
take $l_{1}$ satisfying $l_{1} > g(g-1)$, the pole smaller than the
VEV does not appear in the whole $k_{1}-g$ parameter space. The result
for smaller $\eta$ with $\kappa = 0.0$ is depicted in Fig.
\ref{fig:t_l1_kappa0.0}. We find the critical value of the breaking
scale, $\eta_{cr} \simeq 0.95~(1.00)$ for $\kappa = 0.0~(0.1)$, which
corresponds to $\la \phi \ra \simeq 1/\sqrt{2}$.

%%%%%%%%%%%%%%%%%%%%%%%%%%%%%%%%%%%%%%%%%%%%%%%%%%%%%%%%%%%%%%%%%%%%%%
\section{Conclusions and discussions}
%%%%%%%%%%%%%%%%%%%%%%%%%%%%%%%%%%%%%%%%%%%%%%%%%%%%%%%%%%%%%%%%%%%%%%

We have studied a topological inflation in supergravity. First, we
have shown that topological inflation really takes place in
supergravity. Also, the criterion of successful topological inflation
depends not only on the breaking scale of the discrete symmetry but
also on the mass of the inflaton near the origin. This is because the
inflaton rolls down rapidly from the origin if its mass is large. For
a very flat case favored by the observation of the spectral index,
$n_{s} \simeq 1 - 0.8$ (i.e., $0 < \kappa < 0.1$), we have found that
the critical breaking scale $\eta_{cr}$ becomes as small as $M_{G}$,
which is smaller than the critical value, $\eta_{cr}\cong1.7M_{G}$
observed in Ref.~\cite{sakai}. Finally we have discussed the
primordial spectrum produced by the topological inflation. In general,
the topological inflation predicts the tilted spectrum $n_{s} < 1$
depending on $\kappa$.\footnote{It is possible to produce more exotic
spectrum including blue one. This is due to the exponential blow of
the potential, which is significant for the region $\varphi \gtilde
M_{G}$ and makes the potential more complex than the simple
double-well potential.}

The present topological inflation model is free from the thermal and
nonthermal overproduction of gravitinos since the reheating
temperature can be as low as $10^{8}$~GeV. Furthermore, as pointed out
in Ref.~\cite{Asaka}, this model is consistent with a leptogenesis
scenario in which heavy Majorana neutrinos are produced in the
inflaton decay and successive decays of the Majorana neutrinos result
in lepton asymmetry enough to explain the observed baryon asymmetry in
the present universe.

%%%%%%%%%%%%%%%%%%%%%%%%%%%%%%%%%%%%%%%%%%%%%%%%%%%%%%%%%%%%%%%%%%%%%%
\subsection*{ACKNOWLEDGMENTS}
%%%%%%%%%%%%%%%%%%%%%%%%%%%%%%%%%%%%%%%%%%%%%%%%%%%%%%%%%%%%%%%%%%%%%%

M.Y. is grateful to T. Kanazawa and J. Yokoyama for useful
discussions.  M.K. and T.Y. are supported in part by the Grant-in-Aid,
Priority Area ``Supersymmetry and Unified Theory of Elementary
Particles''(No. 707).  N.S. and M.Y. are partially supported by the
Japanese Society for the Promotion of Science.

\bef
  \bec
    \leavevmode\psfig{figure=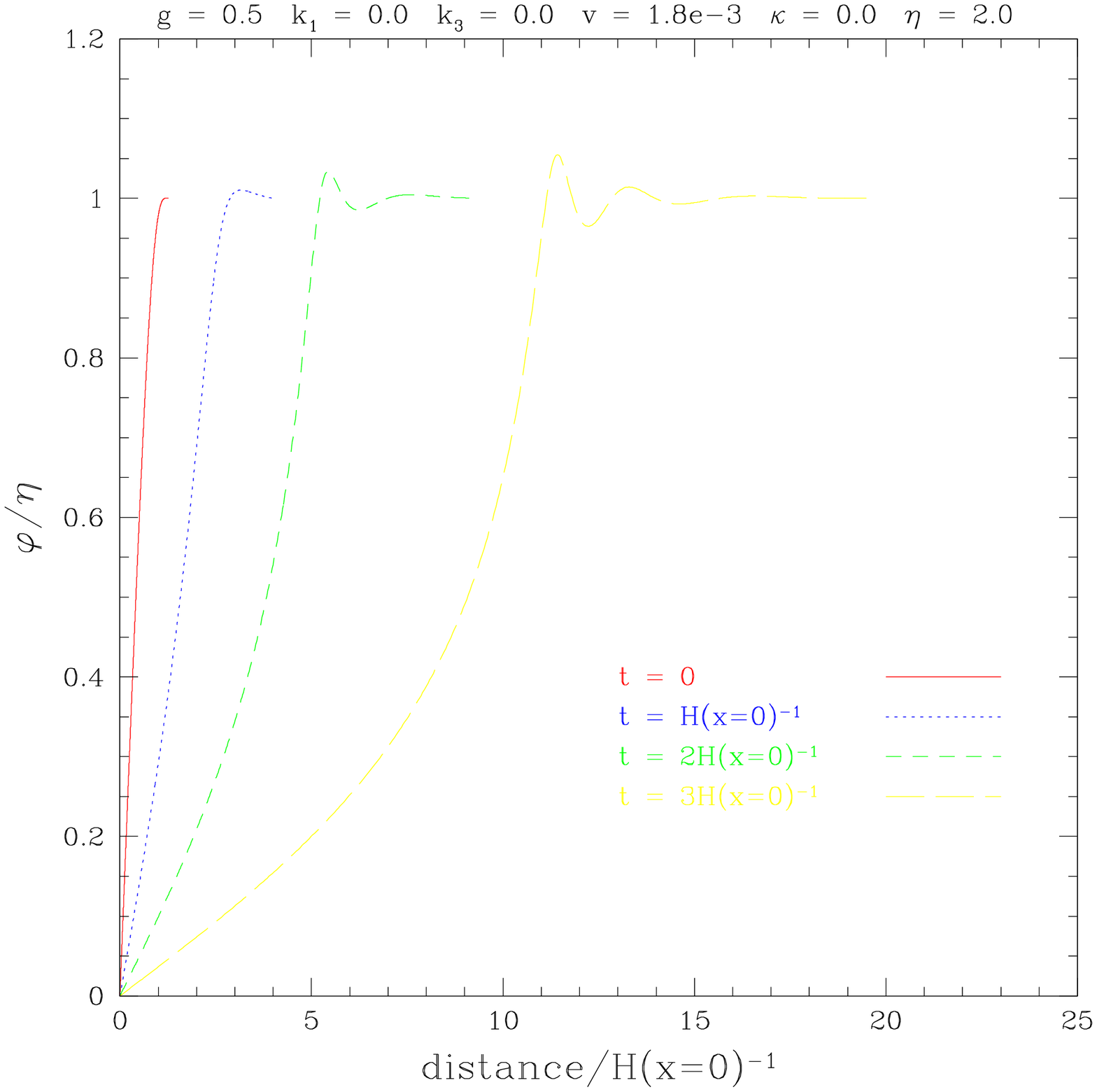,width=16cm}
  \eec
  \caption{Time evolution of the domain wall in the case of $k_{1} =
  0$ and $g = 0.5~(\eta = 2.0)$. The vertical axis represents the
  value of the scalar field. The horizontal axis represents the proper
  distance from the domain wall core. Note that the proper width of
  the wall becomes much larger than the Horizon scale. As time
  elapses, the domain wall expands and topological inflation takes
  place. Here we set $v=1.8 \times 10^{-3}$. But the result does not
  depend on the energy scale $v$ as shown later.}
  \label{fig:minimal}
\eef

\bef
  \bec
    \leavevmode\psfig{figure=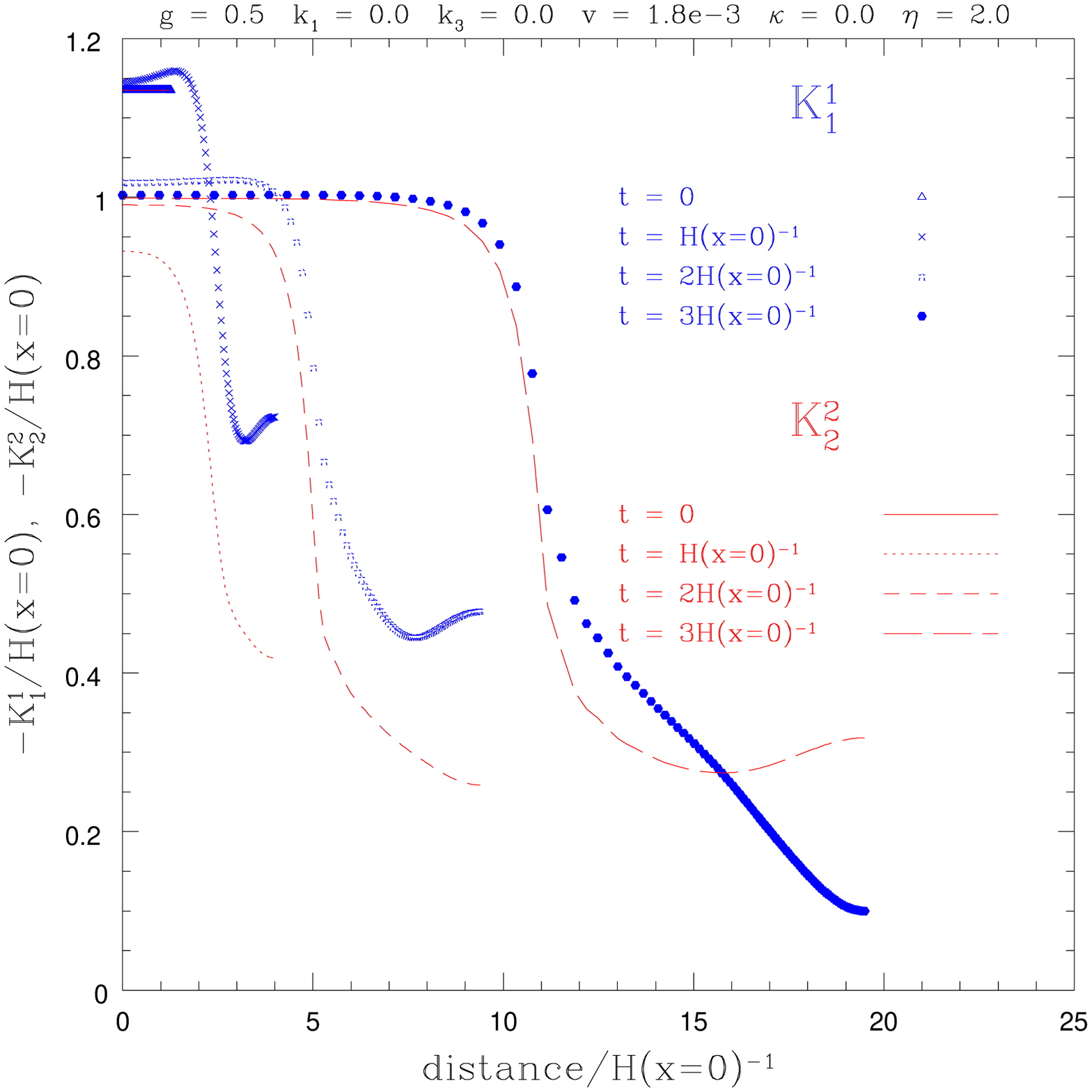,width=16cm}
  \eec
  \caption{The expansion rates of the scale factors, $A$ and $B$, that
  is, $-\CK^{1}_{1} = \dot{A}/A$ and $-\CK^{2}_{2} = \dot{B}/B$ are
  shown for the case of Fig. \ref{fig:minimal}. As the universe
  expands enough, $-\CK^{1}_{1}$ and $-\CK^{2}_{2}$ approach the same
  value, the Hubble parameter $H(x=0)$, inside the wall.}
  \label{fig:tk1k2}
\eef

\bef
  \bec
    \leavevmode\psfig{figure=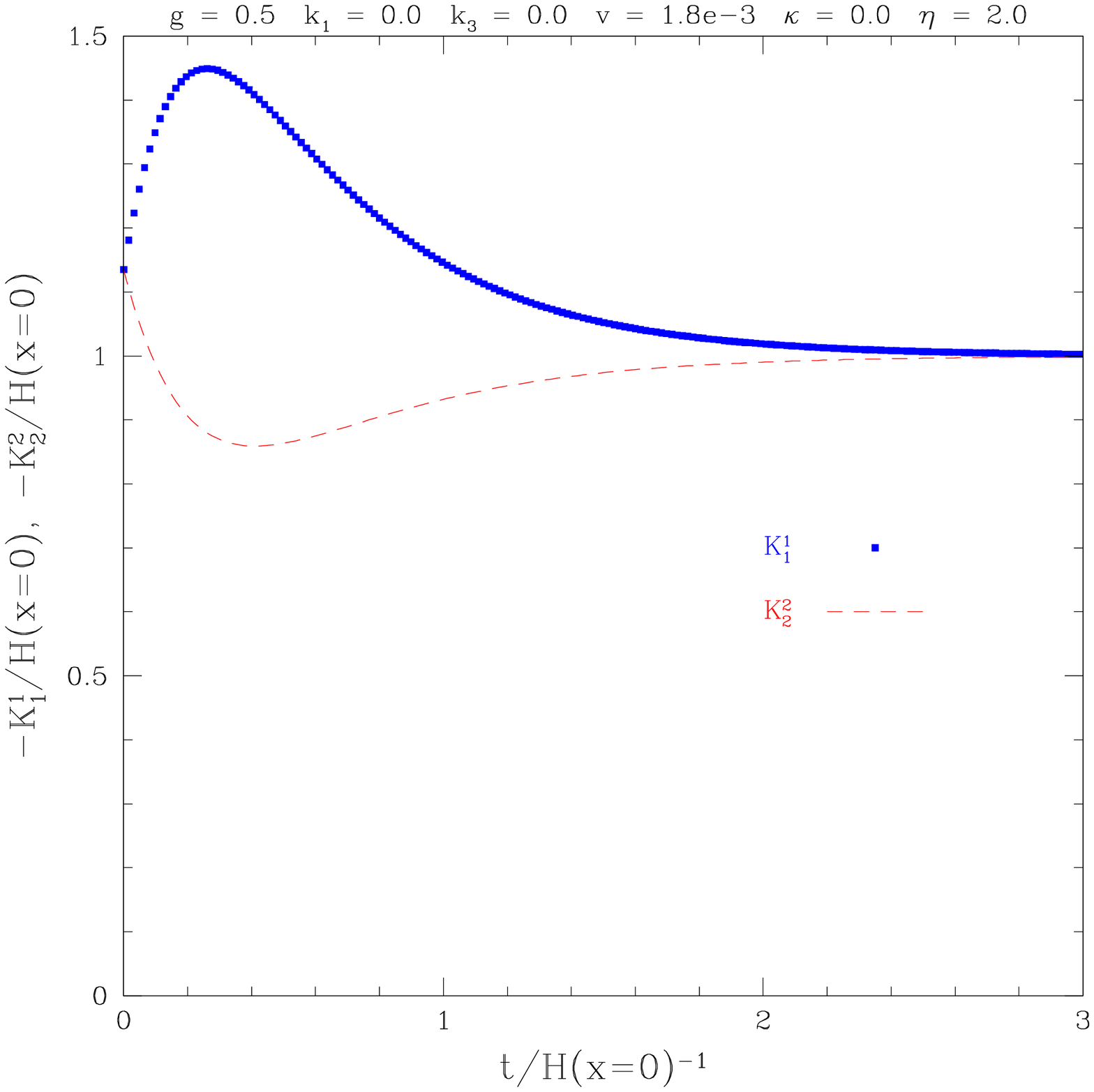,width=16cm}
  \eec
  \caption{Time evolution of $-\CK^{1}_{1}$ and $-\CK^{2}_{2}$ at
  the origin $x=0$ is shown for the case of Fig. \ref{fig:minimal}}
  \label{fig:k1k2x0}
\eef

\bef
  \bec
    \leavevmode\psfig{figure=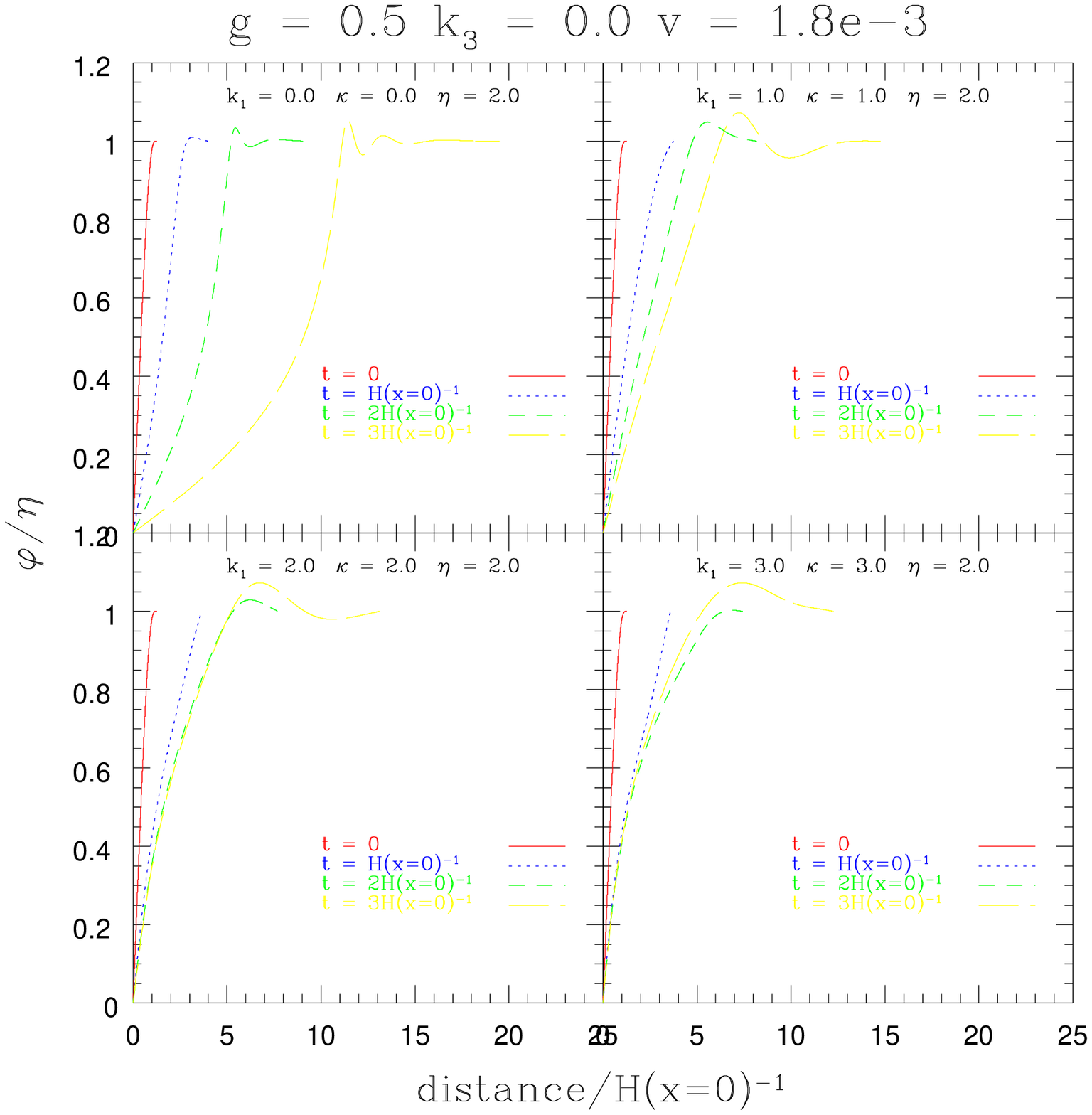,width=16cm}
  \eec 
  \caption{Time evolution of the domain wall in the cases of
  different $\kappa$ for the fixed VEV, $\eta = 2.0~(g = 0.5)$, with
  $v=1.8 \times 10^{-3}$.}
  \label{fig:g=0.5v=1.8-3}
\eef

\bef
  \bec
    \leavevmode\psfig{figure=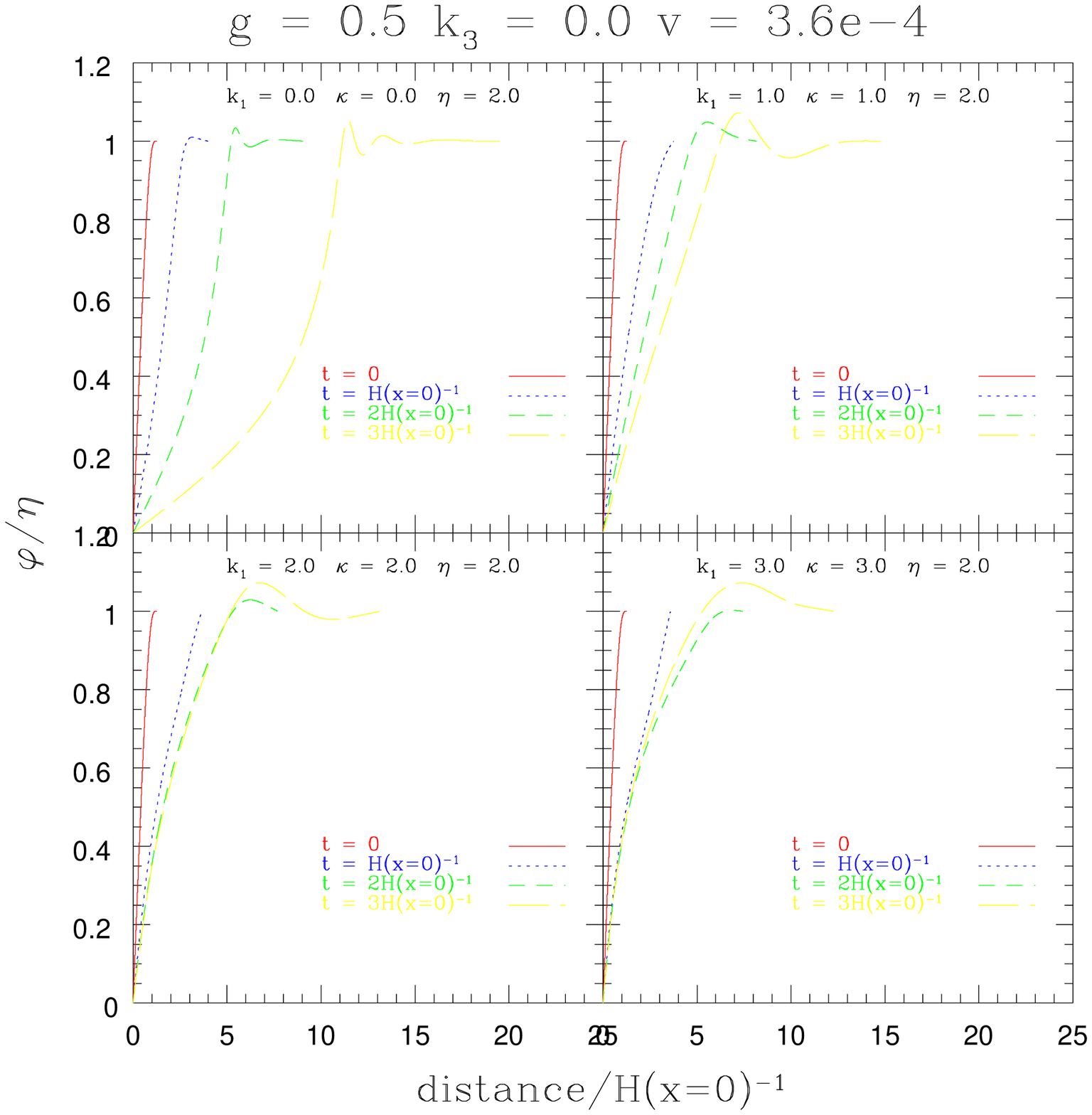,width=16cm}
  \eec 
  \caption{The results with $v = 3.6 \times 10^{-4}$. The energy scale
  $v$ is different from that in Fig. \ref{fig:g=0.5v=1.8-3}.}
  \label{fig:g=0.5v=3.6-4}
\eef

\bef
  \bec
    \leavevmode\psfig{figure=t_k3.epsi,width=16cm}
  \eec 
  \caption{The results for the same parameters in Fig.
  \ref{fig:minimal} except $k_{3}=\pm 0.3$.}
  \label{fig:t_k3}
\eef

\bef
  \bec
    \leavevmode\psfig{figure=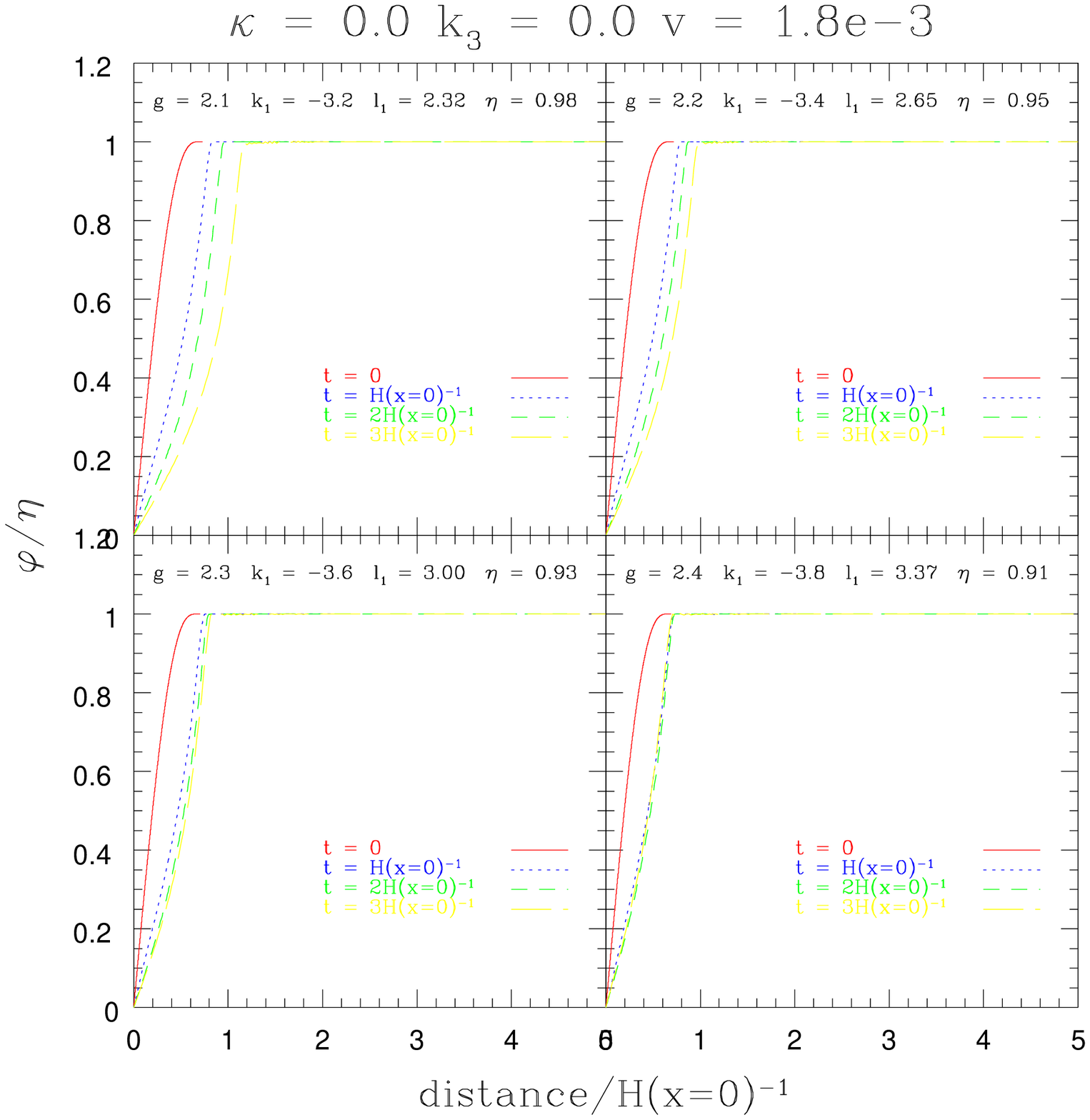,width=16cm}
  \eec 
  \caption{The results with $\eta \sim \eta_{cr}$ for $\kappa = 0.0$.
  They show $\eta_{cr} \simeq 0.95$ for $\kappa = 0.0$}
  \label{fig:t_l1_kappa0.0}
\eef

\end{document}